\documentclass{aa}

\usepackage{txfonts}
\usepackage{threeparttable}
\usepackage{natbib}
\usepackage{kantlipsum}
\usepackage{amsmath}
\usepackage{subfloat}
\usepackage{subfig}
\usepackage{multicol}
\usepackage{graphicx}
\usepackage{array}
\usepackage{color}
\usepackage{url}
\usepackage{enumitem}
\usepackage{dblfloatfix} 
\usepackage{bm}
\bibpunct{(}{)}{;}{a}{}{,}
\providecommand{\kms}{\,km\,s$^{-1}$}
\newcommand{\teff}{T_{\mbox{\textrm{\scriptsize eff}}}}
\newcommand{\teffprime}{T^{\prime}_{\mbox{\textrm{\scriptsize eff}}}}
\newcommand{\gk}{G-K_{\mbox{\textrm{\scriptsize S}}}}
\newcommand{\jk}{J-K_{\mbox{\textrm{\scriptsize S}}}}
\newcommand{\ks}{{K}_{\mbox{\textrm{\scriptsize S}}}}
\newcommand{\tnorm}{\hat{T}}
\newcommand{\Egk}{{\mbox{E}(G-K_{\mbox{\textrm{\scriptsize S}}})}}
\newcommand{\Ejk}{{\mbox{E}(J-K_{\mbox{\textrm{\scriptsize S}}})}}
\newcommand{\feh}{\textrm{[Fe/H]}} 

\begin{document}

   \title{The empirical $Gaia~ G$-band extinction coefficient}

   \author{
   	C. Danielski  \inst{1}
		\and
         C. Babusiaux \inst{1,2}, L. Ruiz-Dern \inst{\inst1}, P. Sartoretti \inst{1}, F. Arenou\inst{1}
          }

   \institute{GEPI, Observatoire de Paris, PSL Research University, CNRS, 5 Place Jules Janssen, 92190, Meudon, France \\
   \email{camilla.danielski@obspm.fr}
   \and
   Univ. Grenoble Alpes, CNRS, IPAG, 38000 Grenoble, France
}

  \abstract
  {The first $Gaia$ data release unlocked the access to the photometric information of 1.1 billion sources in the $G$-band. 
   Yet, given the high level of degeneracy between extinction and spectral energy distribution for large passbands such as the $Gaia~ G$-band, a correction for the interstellar reddening is needed in order to exploit $Gaia$ data.}
   {The purpose of this manuscript is to provide the empirical estimation of the $Gaia$~$G$-band extinction coefficient k$_{G}$  
   for both the red giants and main sequence stars, in order to be able to exploit the first data release DR1. }
   {We selected two samples of single stars: one for the red giants  and one for the 
   main sequence.  Both samples are the result of a cross-match between Gaia DR1 and 2MASS catalogues; they consist of high quality photometry in the $G$-, $J$- and $\ks$-bands. These samples were complemented  by temperature and metallicity information retrieved from, respectively, APOGEE DR13 and LAMOST DR2 surveys.  
   We implemented a Markov Chain Monte Carlo method where we used  $(\gk)_\mathrm{0}$ vs $\teff$ and $(\jk)_\mathrm{0}$ vs $(\gk)_\mathrm{0}$
calibration relations to estimate the extinction coefficient k$_{G}$ and we quantify its corresponding confidence interval via bootstrap resampling method.
We tested our method on samples of red giants and main sequence stars, finding consistent solutions.}  
   {We present here the  determination of the $Gaia$ extinction coefficient through a completely empirical method. Furthermore we provide the scientific community a formula for measuring the extinction coefficient as a function of stellar effective temperature, the intrinsic colour  $(\gk)_\mathrm{0}$ and absorption.}
   {}


 \authorrunning{C. Danielski}
   \maketitle
%

\begin{figure*}[h!]
\centering
\subfloat[APOGEE stars - Data Release 13.]{\label{fig:apogee_sample}\includegraphics[width=0.42\linewidth]{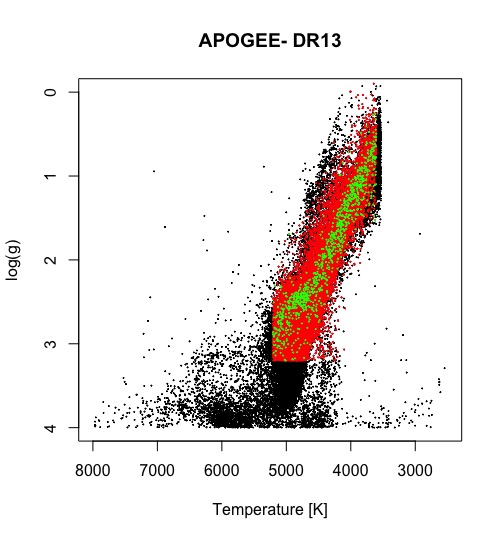}} \qquad \qquad
\subfloat[LAMOST stars - Data Release 2.]{\label{fig:lamost_sample}\includegraphics[width=0.42 \linewidth]{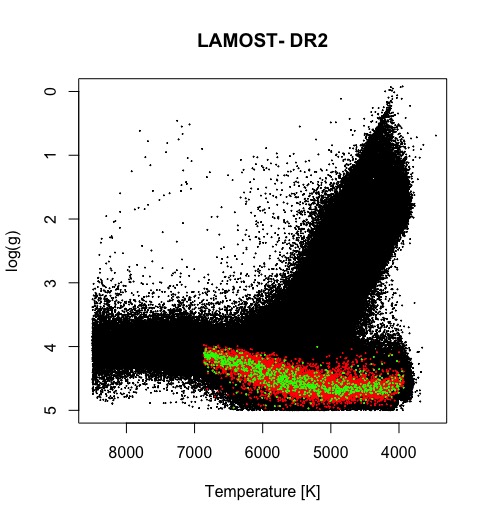}}
\caption{Data selection: spectroscopic surveys used for the red giants (\textit{left}) and the main sequence (\textit{right}) analysis. Black points represent stars in each data release, red points are objects in the first sample selection, used to differentiate red giants stars from dwarfs and viceversa.  Green points represent
the 1000 stars temperature-sampling selection we have used to measure the extinction coefficient.}
\label{fig:Data}
\end{figure*}

\section{Introduction}

When it comes to understanding the physics of disk galaxies, our location within the Milky Way plays an important role. 
By observing our visible sky and studying the astrophysical processes of its individual components, we can learn about the structure and dynamics of the Galaxy, and hence infer its formation and evolution. This prospect would not be possible only by examining other galaxies.\\
Accordingly, numerous spectro/photometric surveys have been conducted over the last decade, altogether spanning different spectral ranges to cover a wide variety of galactic astrophysical processes. 
To mention some: the Fermi Gamma-ray space Telescope (GLAST, \citealt{GLASTref}) in the gamma-ray range,  XMM-Newton (\citealt{XMMref}, 
\citealt{XMMsurvey}) in the X-ray,  the Galaxy Evolution Explorer (GALEX, \citealt{GALEXref}) in the ultraviolet (UV), the Sloan Digital Sky Survey 
(SDSS, \citealt{SDSSref}) in the optical,  the 2-micron All-Sky Survey (2MASS, \citealt{2MASSref}) in the near infrared (NIR) and Planck \citep{Planck} in the far infrared-microwave range. 

Yet among all, the mapping process of the Milky Way is culminating with $Gaia$, the ESA space mission that has just started providing data to study formation, dynamical, chemical and star-formation evolution \citep{Perryman2001, GaiaRelease}.
Nonetheless, despite the unrivalled completeness of its information, $Gaia$, like the other surveys, does not rule out astrophysical selection effects  such as the interstellar extinction.

Extinction is caused by the presence of dust in the line of sight and it has the main effect of dimming sources and reddening them. 
In particular, around 30$\%$ of light in the  UV, optical and NIR is scattered and absorbed due to the interstellar medium \citep{Draine2003}.
In broad-band photometry, additionally, a major hurdle to face is the substantial degeneracy between extinction, effective temperature $\teff$ and 
spectral energy distribution (SED).
This degeneracy limits the accuracy by which any of the parameters can be estimated \citep{BailerJ2010b}. 
Important to mention that extinction coefficients k$_{\lambda}$  are a function of wavelength and get greater towards shorter wavelengths; they are defined as k$_{\lambda}$= $A_{\lambda}/A_\mathrm{ref}$ where $A_{\lambda}$ is the absolute absorption at any wavelength, expressed relative to the absolute absorption at a chosen reference wavelength $A_\mathrm{ref}$ \citep{Cardelli89}.\\
Over recent years an increased number of studies focused on delivering more precise extinction coefficients values for various known pass-bands by using a combination of spectroscopic and photometric information retrieved from the most advanced surveys (e.g. \citealt{Yuan2013}, \citealt{Schlafly2016}, \citealt{Xue2016}).\\
Important to note though that in case of a wide pass-band, like the $Gaia$ one, a star which has the greater fraction of its radiation in the blue-end of the spectrum (a bluer star), has a larger extinction coefficient than a redder star.
It is hence mandatory to have exact knowledge of the passband to correctly estimate the reddening factor.

Reddening of an object in a given colour can be described  by the  \textit{colour excess} which is the difference between its observed colour 
and its intrinsic value. For instance the colour excess between the $Gaia$ $G$-band and the 2MASS $\ks$-band is given by $\Egk$  = ($\gk)_\mathrm{obs} - (\gk)_\mathrm{0}$ 
where ($\gk)_\mathrm{obs}$ is the observed colour
and ($\gk)_\mathrm{0}$  is the intrinsic one.

At the time of the publication of $Gaia$ DR1, only the nominal $Gaia~G$-passband, modelled with the most up-to-date pre-launch information, was available \citep{Jordi2010}. Recently a calibration of the $Gaia~G$-DR1 passband has been provided by \cite{Maiz2017}. The second is redder than the first one due to some water contamination in the optics, which diminished the spectral efficiency more in the blue part of the band than in the red one \citep{GaiaRelease}.
A new filter response curve will be available with the second $Gaia$ data release (DR2).
Uncertainties, either in the passband determination or in the extinction law or in the stellar model atmospheres, can yield to inaccurate extinction coefficients.
For these reasons and because the accurate determination of reddening to a star is key for exploiting the available $Gaia$ data,
we present here a determination of $Gaia$ extinction coefficient for both red giants and dwarfs stars through a completely empirical method.\\
~\\
\indent The manuscript is structured as follows.
\S \ref{sec:data} introduces the data we used and describes the data selection for the red giants and dwarfs sample respectively.
In \S \ref{sec:calibration} we estimate the photometric calibration relations for the main sequence.
\S \ref{sec:ASPOS} describes the estimation of the theoretical extinction coefficients used in our analysis.
In \S  \ref{sec:method} we present the technique we used to estimate $Gaia$ $G$-band extinction coefficient (k$_{G}$) for the red giants sample, the dwarfs sample and, finally, for the union of both samples.  
In \S \ref{sec:results} we present the results and discuss them. Finally, \S \ref{sec:conclusions} presents our conclusions.

\begin{figure*}[b!!]
\centering
	\includegraphics[width=0.40\linewidth]{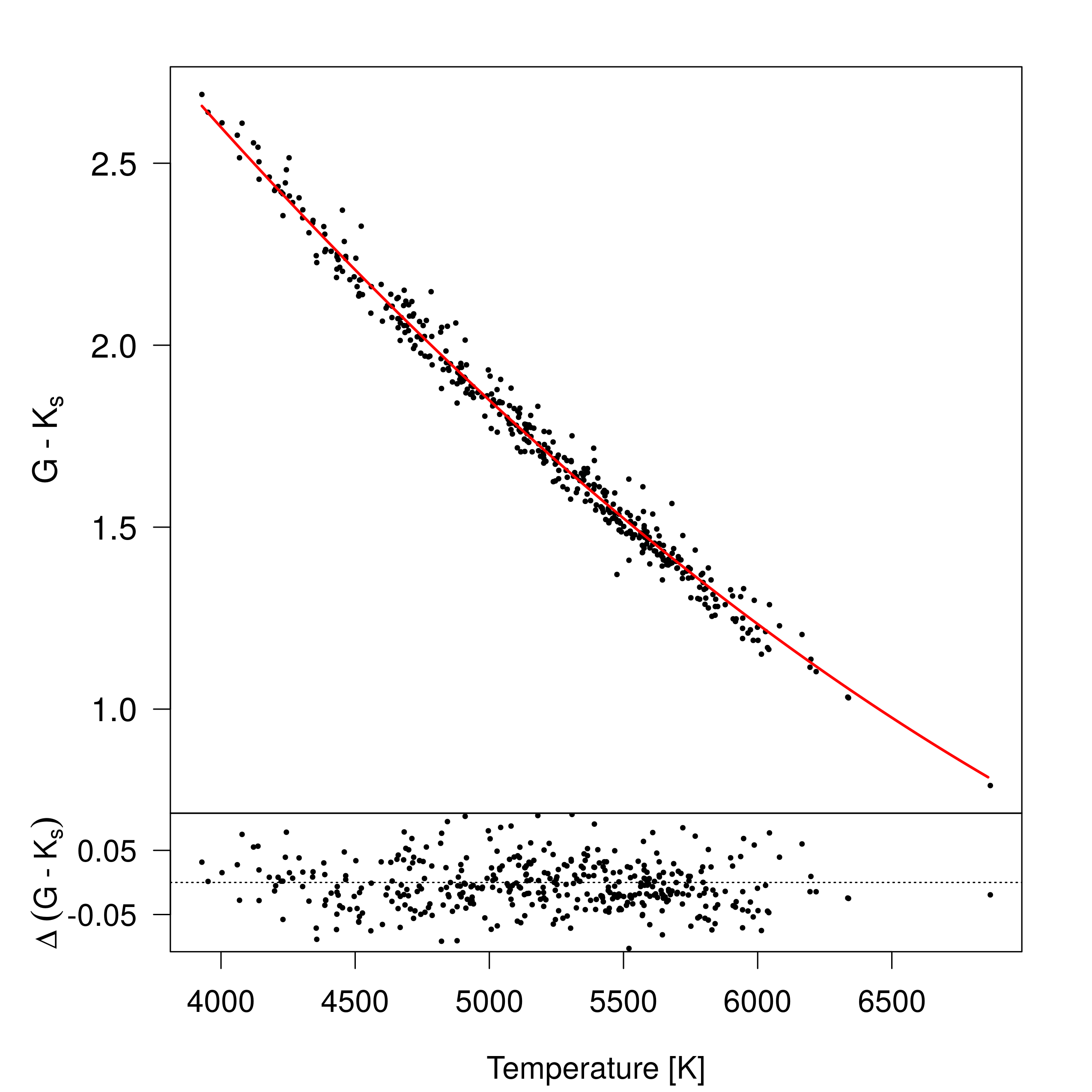} \qquad \qquad
	\includegraphics[width=0.4\linewidth]{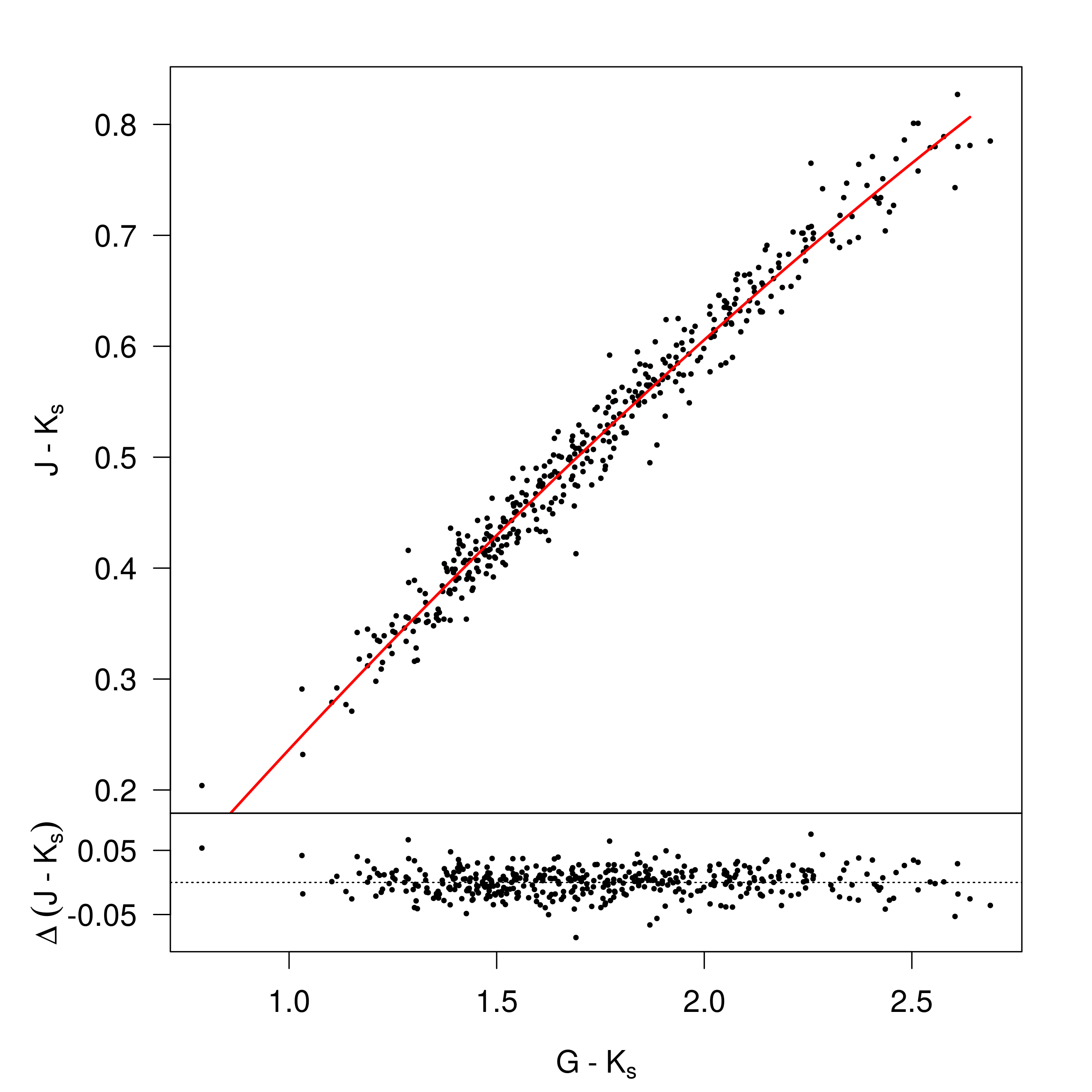}
\caption{Main sequence photometric calibration relations: ($\gk)_{0}$ as a function of $\teff$ (\textit{left}); ($\jk)_{0}$ as a function of ($\gk)_{0}$, (\textit{right}). \textit{Top panels}: stars (black dots) in the sample where solid red lines correspond to our calibration at solar metallicity (Eq. \ref{eq:calibration}). \textit{Bottom panels}: residuals (black dots) corresponding to the top panel calibration.}
\label{fig:lamostcalib}
\end{figure*}

\section{Data}
\label{sec:data}
For our analysis we cross-matched photometric  and spectroscopic data from different surveys.
More specifically for the photometric information we used the $Gaia$ DR1 and 2MASS catalogues. 
The 2MASS $J$, $H$, $\ks$ magnitudes are available for a large fraction of the $Gaia$ sources and the near-infrared extinction law is fairly well characterized (e.g. \citealt{FitzpatrickMassa2009}). \\
Spectroscopic parameters, such as effective temperature $\teff$, surface gravity $\log(g)$ and metallicity $\feh$, were retrieved from surveys selected $ad~ hoc$ for the samples analysed.
Our analysis was performed on both the red giants  (RG) sample and the main sequence (MS) one (Fig. \ref{fig:Data}) separately, then on both samples combined together.

\subsection{The Red Giant sample}
\label{sec:redclumpdata}
Effective temperature $\teff$, surface gravity $\log(g)$ and metallicity $\feh$ were retrieved from the spectroscopic survey APO Galactic Evolution Experiment  (APOGEE), DR13 \citep{ApogeeDR13}.\\
The cross-match between APOGEE and $Gaia$ was done using the 2MASS ID provided in APOGEE and the 2MASS-GDR1 cross-matched catalogue \citep{Marrese2017}, where we kept only cross-matched stars with angular distance lower than 0.3\arcsec.
Hence, we selected those stars with high infrared photometric quality (i.e. 2MASS "AAA" quality flag),  radial velocity error 
$\sigma_\mathrm{RV}$ < 0.1 \kms\ to exclude binary stars, and photometric errors  of $\sigma_{G}$ < 0.01~mag,   
$\sigma_{J}$ < 0.03~mag and $\sigma_{\ks}$ < 0.03~mag. The $G$-band photometric error has been later increased of 0.01~mag in quadrature to mitigate the impact of bright stars residual systematics  (\citealt{Evans2017}, \citealt{Arenou2017}).
Then we retained the red giants stars by screening those with colour ($\gk)_\mathrm{obs}$ > 1.6~mag. For stars with parallax information in $Gaia$ DR1 (TGAS), we used the same criteria as \cite{Laura17}:
\begin{equation*}
\begin{array}{cc}
G + 5+ 5~\mbox{log}_{10} \left( \frac{\varpi + 2.32 \sigma_{\varpi}}{1000}\right) < 4
\end{array}
\end{equation*}

\noindent where the factor 2.32 on the parallax error $\sigma_{\varpi}$ corresponds to the 99th percentile of the 
parallax probability density function. 
When no parallax information was available, the selection was performed by 
filtering on the surface gravity ($\log(g)$ < 3.2~dex).\\
Finally, we selected those stars with effective temperature 3603 K  < $\teff \pm \sigma_\mathrm{Teff}$ < 5207 K and metallicity -1.5~dex < $\feh$ < 0.4~dex.
to work within the same  limits set for the $\teff$ vs ($\gk)_\mathrm{0}$
calibration  by \cite{Laura17}.
The application of these criteria delivered a sample of 71290 stars.

\begin{table*}[t!]
\centering
\begin{tabular}{c|c|cccccc }
\multicolumn{8}{c}{PHOTOMETRIC CALIBRATION COEFFICIENTS}\\
\hline\hline
 & RMS & $c_1$ & $c_2$ & $c_3$ & $c_4$ & $c_5$ & $c_6$ \\
\hline
RG & 0.05 &13.554 $\pm$ 0.478 & -20.429$\pm$1.020  &  8.719$\pm$0.545 & 0.143$\pm$0.013 & -0.0002$\pm$0.009 & --   \\
MS & 0.04 &6.946$\pm$0.181 & -6.835$\pm$0.354  &  1.711$\pm$0.172  & -- & -- & -- \\
\hline
& &$c_7$ & $c_8$ & $c_9$ & $c_{10}$ & $c_{11}$ & c$_{12}$ \\
\hline
RG & 0.02 & -0.227$\pm$ 0.024 & 0.466$\pm$0.021  & -0.023$\pm$0.005 &-0.016$\pm$0.002 &-0.005$\pm$0.001& --  \\
MS & 0.02 &-0.200$\pm$0.034 & 0.471$\pm$0.038 & -0.03$\pm$0.01 & --& -- & -- \\
\hline
\end{tabular}
\caption{Coefficients of the calibration relations (Eq. \ref{eq:calibration}) and their uncertainties for the RG sample \citep{Laura17} and MS sample (this work). The RMS corresponds to one standard deviation of the relations residuals.}
\label{tab:CalibParams}
\end{table*}

\subsection{The main sequence sample}
\label{sec:dwarfs}

For the dwarfs sample we cross-matched our photometric samples with the 
Large sky Area Multi-Object fiber Spectroscopic Telescope  survey (LAMOST, \citealt{LAMOST}) DR2, 
from which we retrieved effective temperature $\teff$, surface gravity $\log(g)$ and metallicity $\feh$.
The cross-match with 2MASS and $Gaia$ DR1 was done with a radius of 0.2\arcsec.
We selected a sub-sample of objects with radial velocity error 
$\sigma_\mathrm{RV}$ < 20 \kms\ to exclude binary stars, photometric errors  $\sigma_\mathrm{G}$,
$\sigma_\mathrm{K_S}$, $\sigma_\mathrm{J}$  < 0.03~mag and relative temperature error smaller than 5$\% $.
As explained in \S \ref{sec:redclumpdata}, we increased $\sigma_\mathrm{G}$ of 0.01~mag in quadrature.
Following we retained the main sequence stars by applying both colour and surface gravity cuts:

\begin{itemize}
\item[] $\log(g)$ -2$\sigma_{\log(g)}$ > 3.5~dex
\end{itemize}

\noindent where $\sigma_{\log(g)}$ is the surface gravity error.\\
We set the metallicity range for the MS calibration (and consequently for the extinction coefficient estimation) to be solar-like ( -0.05~dex < $\feh$ < 0.05~dex) because of the significant correlation between metallicity and effective temperature in the LAMOST data which did not allow a good convergence of the photometric calibration (see \S \ref{sec:calibration}).

We further selected stars with temperature within the calibration temperature interval (3928 K < $\teff \pm \sigma_\mathrm{Teff}$ < 6866 K), leaving a final sample of 17468 dwarfs.

\begin{figure*}[b!]
 \centering
 \includegraphics[width=0.3\linewidth, trim = 30 10 30 30]{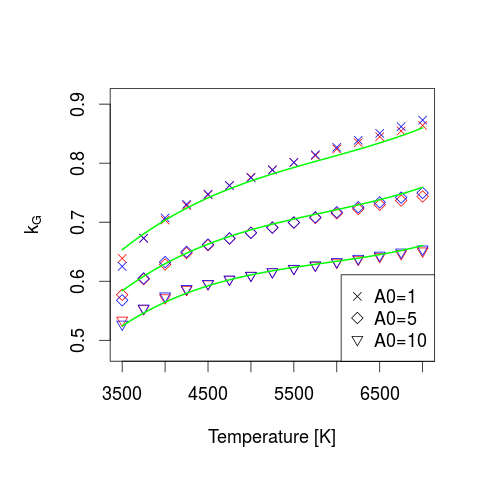}\quad
 \includegraphics[width=0.3\linewidth, trim = 30 10 30 30]{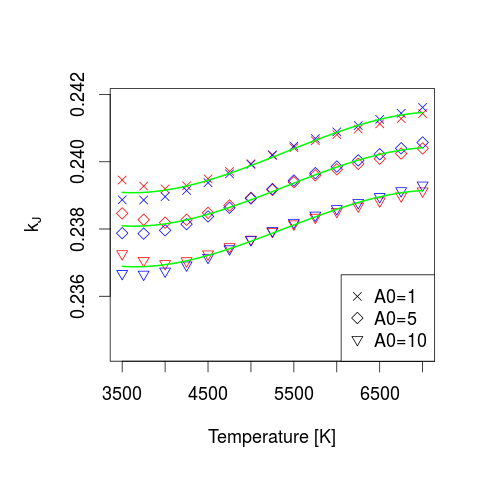}\quad
 \includegraphics[width=0.3\linewidth, trim = 30 10 30 30]{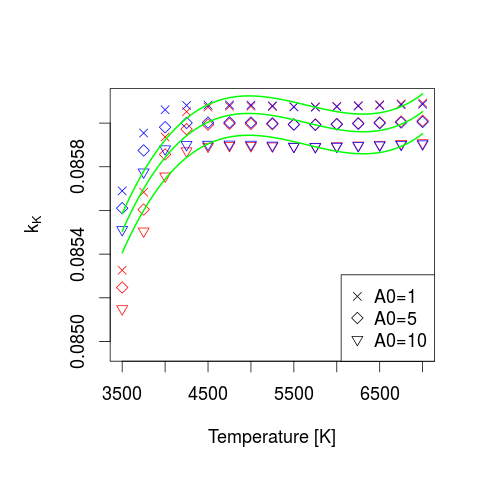}
 \caption{Theoretical extinction coefficients in the $G$- ($Gaia$), $J$- and $\ks$ (2MASS) bands as a function of temperature for different extinctions ($A_0$ = 1, 5, 10 mag) and different surface gravities: $\log(g)$ = 2.5 dex (\textit{red}) and $\log(g)$ = 4 dex (\textit{blue}). Green lines represent the global fit for the three absorption values.}
 \label{fig:theoreticalcoefs}
\end{figure*}

\section{Photometric Calibration} 
\label{sec:calibration}

In order to empirically measure the $Gaia$ $G$-band extinction coefficient k$_{G}$, the colour excess $\Egk$ and $\Ejk$ for our samples need to be determined.
To do so  we used for the RG sample the photometric calibration relations presented in \cite{Laura17} while, for the MS sample, we applied the method described therein to empirically retrieve the photometric 
calibration relations.
Specifically, the calibration relations for both samples were modelled as the following: 
~
\begin{equation}
\begin{array}{ccl}
(\gk)_\mathrm{0} &= &c_1 + c_2 ~\tnorm + c_3 ~\tnorm^2 + c_4~\feh+...\\
& & + c_5~\feh^2 +~c_6~\tnorm~\feh^2\\
(\jk)_\mathrm{0} &= &c_7 + c_8~(\gk)_\mathrm{0} + c_9~(\gk)_\mathrm{0}^2 ~+...\\
 & & +~ c_{10}~\feh + c_{11}~\feh^2 ~+~...\\
 & & c_{12}~(\gk)_\mathrm{0}~\feh~
\end{array}
\label{eq:calibration}
\end{equation}
~

\noindent where $\tnorm = \teff$/5040 is the normalised temperature and $c_i$ are the coefficients reported in Table \ref{tab:CalibParams} for both RG and MS samples.

For calibrating the main sequence relations we selected from the sample of \S\ref{sec:dwarfs} only low extinction stars (E($B-V$) $<$ 0.01) selected from the recent 3D local extinction map of \cite{Capitanio2017} or the 2D map of \cite{Schlegel98} when no distance information was available. We required the relative temperature error to be smaller than  2$\% $. 
The application of these further criteria left a total of 415 stars that we used for the calibration process. Please  refer to \cite{Laura17} for more details on the calibration method. 
Fig. \ref{fig:lamostcalib} shows the relations (Eq. \ref{eq:calibration}, Table \ref{tab:CalibParams}) which were established within the interval of temperature [3928 K, 6866 K].  

\begin{table*}[t!]
 \centering
\fontsize{9}{13}\selectfont
\resizebox{\textwidth}{!}{%
\begin{tabular}{c|ccccccc}
\multicolumn{8}{c}{THEORETICAL EXTINCTION COEFFICIENTS}\\
 \hline \hline
  RG + MS & $a_{1}$ & $a_{2}$& $a_{3}$ & $a_{4} $&  $a_{5} $& $a_{6}$ & $a_{7}$\\
 \hline 
k$_{G}~(\tnorm, A_0)$ & -0.317797257 &	2.538901003	&-1.997742387&	0.572289388	&-0.013179503	&0.000607315	&-0.01126344 \\
k$_{G}~(\tnorm, A_0)$ - DR1 &  -0.279133556	 & 2.373624663	 & -1.878795709 & 0.53904796 & -0.011673326 & 0.000544945 & 	-0.010233727\\
k$_{J}~(\tnorm, A_0)$ & 0.252033852	&-0.042526876&	0.044560182	&-0.013883035&	-0.000239872&	8.45E-07&	-1.96E-05 \\
k$_{\ks}~(\tnorm, A_0)$ & 0.073839341&	0.03381806&	-0.03063728&	0.009124118&	-1.95E-05&	9.66E-09	&-6.81E-07\\
  \hline 
  k$_{G}~((\gk)_\mathrm{0}, A_0)$ & 0.935556283	&-0.090722012&	0.014422056	&-0.002659072&	-0.030029634	&0.000607315	&0.002713748 \\
k$_{J}~((\gk)_\mathrm{0}, A_0)$ & 0.242998063	&-0.001759252&	0.000107601	&2.54E-05&	-0.000268996	&8.45E-07	& 4.60E-06 \\
k$_{\ks}~((\gk)_\mathrm{0}, A_0)$ & 0.086033161&	7.65E-05&	8.54E-06&	-1.52E-05	&-2.00E-05&	- & 	- \\
  \hline 
k$_{G}~((\gk)_\mathrm{0}, A_0)$ - DR1 & 0.882095056 & -0.086780236 & 0.01511573 & -0.002963829 & -0.027054718 & 0.000544945 & 0.002604135 \\
k$_{J}~((\gk)_\mathrm{0}, A_0)$ - DR1 & 0.243062354 & -0.001899476 & 0.000140615 & 2.58E-05 & -0.000269124 & 8.45E-07 & 4.86E-06 \\
k$_{\ks}~((\gk)_\mathrm{0}, A_0)$ - DR1 & 0.086025432 & 9.23E-05 & 3.32E-06 & -1.65E-05 & -2.00E-05 & - & -  \\
\hline

 \end{tabular}
   }
 \caption{Theoretical extinction coefficients for the $Gaia ~G$-band, both pre-launch \citep{Jordi2010} and \textit{G}-DR1  \citep{Maiz2017} passbands, and for $J$- and $\ks$-bands, measured by using the \cite{FitzpatrickMassa} extinction law and the Kurucz Spectral Energy Distributions from \cite{CastelliKurucz03} (see \S \ref{sec:ASPOS}). The extinction coefficients are modelled as function of ($\tnorm, A_{0}$) and ($(\gk)_\mathrm{0}, A_0$), Eq. (\ref{eq:coefformula}), and they are valid for  both the red giants and the main sequence samples for 3500~K$<\teff<$7000~K and $A_0$ < 20 mag}
  \label{tab:theoreticalk}
\end{table*}

\section{Theoretical extinction coefficients}
\label{sec:ASPOS}

We computed the theoretical extinction coefficients k$_m$ using the \cite{FitzpatrickMassa} extinction law $E_\lambda$, the Kurucz Spectral Energy Distributions $F_\lambda$ from \cite{CastelliKurucz03} and the filters transmissions $T_\lambda$ :
\begin{equation}
k_m A_0 = A_m = m - m_0 = -2.5 \log_{10}\left(\frac{\int F_\lambda T_\lambda E_\lambda^{A_0} d\lambda}{\int F_\lambda T_\lambda d\lambda}\right) 
\end{equation} 
with $A_0$ the interstellar extinction at $\mathrm{\lambda}$ = 550 nm ($Gaia$ reference value). 
While the \cite{FitzpatrickMassa} extinction law was derived using hot stars, this extinction law was calibrated using the full star spectral energy distribution and therefore should be also valid for the lower temperature stars of our sample.

For 2MASS transmissions were taken from \cite{Cohen03}\footnote{\url{http://www.ipac.caltech.edu/2mass/releases/allsky/doc/sec6_4a.html}}. 
For comparison purposes we used the $Gaia$ pre-launch transmission\footnote{\url{https://www.cosmos.esa.int/web/gaia/transmissionwithoriginal}} and the \textit{Gaia~G}-DR1 transmission of \cite{Maiz2017}\footnote{\url{http://cdsarc.u-strasbg.fr/viz-bin/qcat?J/A+A/608/L8}}.

As shown by \cite{Jordi2010}, in such a large band as $Gaia$ $G$-band ($\sim$330 - 1050 nm), the extinction coefficient 
varies strongly with temperature and the extinction itself, but less with surface gravity and metallicity. We therefore modelled the extinction coefficients as a function of $A_{0}$ and $\teff$, following the formula:
~
\begin{equation} 
\mbox{k}_m =  a_{1} + a_{2}\mbox{ X} + a_{3}\mbox{ X} ^2 + a_{4}\mbox{X} ^3 + a_{5} A_0 + a_{6}A_0^2
+ a_{7}\mbox{X}A_0
\label{eq:coefformula}
\end{equation}

\noindent where X is either $\tnorm = \teff$/5040 or ($\gk)_\mathrm{0}$, depending if we are analysing the extinction coefficient as a function of the normalised temperature or the colour, respectively. The parameters $a_{i}$ are the coefficients of the fit in each photometric band $m$. 

Table \ref{tab:theoreticalk}  reports the coefficients $a_i$ for the theoretical estimation of the global
k$_{J}$ and k$_{\ks}$ coefficients valid for both red giants and main sequence stars, as well as k$_{G}$, which was computed by using the $Gaia$ pre-launch modelled filter response.

The fit is performed using extinctions computed on a grid with a spacing of 250~K in $\teff$ and 0.01~mag in $A_0$ with 0.01~mag 
$<\mathrm{A}_0<20$ mag and 3500 K $<\teff< 7000$ K and two surfaces gravities: $\log(g)$ = 2.5~dex and 4~dex. The result is shown in Fig. \ref{fig:theoreticalcoefs}.
We checked that high order parameters in the polynomials are needed with an ANOVA test. Only for the $\ks$-band and for relatively low extinctions ($A_0<5$ mag) coefficients $a_5$ and $a_6$ are not significant. 
In particular residuals of the fit are smaller than 0.3\% for k$_{\ks}$, 0.2\% for k$_{J}$ and 4.5\% for k$_{G}$. For k$_{G}$ residuals decrease to 2.4\% when the fit is performed just for $A_0<5$ mag. 
For comparison, we have estimated the extinction coefficients using the  \cite{Cardelli89} extinction law, and compared them to the ones obtained with the \cite{FitzpatrickMassa} law: the difference is of 37\% in the $\ks$-band, 20\% in the $J$-band and 5\% in the nominal $G$-band. 
\cite{Jordi2010} also assessed that $R_V$ variation does not have a significant impact on k$_{G}$. On the other hand, the $\ks$- and $J$-bands are much less sensitive to spectral type variations (Fig. \ref{fig:theoreticalcoefs}): for instance
the difference between a red giant star and Vega is of only 0.07\% for the 
$\ks$-band, 1\% for the $J$-band, while it is of 21\% in the $G$-band.

\begin{figure*}[t!]
\centering
\includegraphics[width=0.3\linewidth, trim = 50 10 20 10]{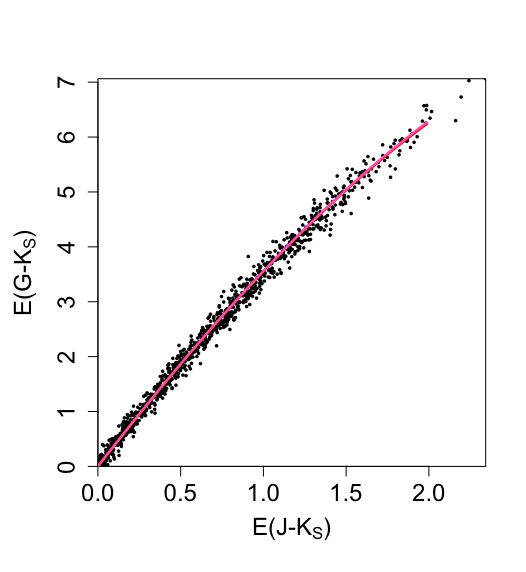} \quad 
\includegraphics[width=0.3\linewidth, trim = 50 10 20 10]{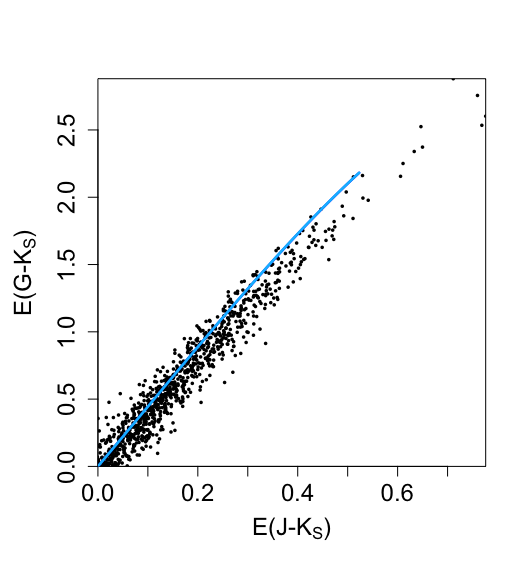} \quad
\includegraphics[width=0.3\linewidth, trim = 50 10 20 10]{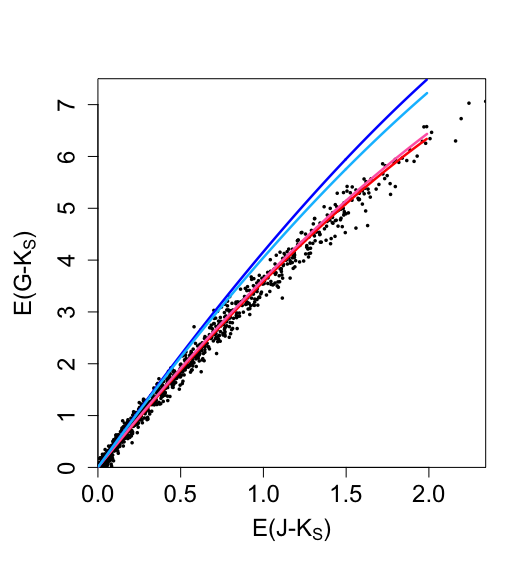} 

\caption{Colour excesses E$(\gk)$ versus E$(\jk)$ for the red giants sample (\textit{left}), the main sequence sample (\textit{centre}) and the combined sample (\textit{right}). Black dots are the 1000 stars selected in each sample. Solid lines represent the colour excess increase with extinction for a reference temperature $\teff$  (4136 K for RG, \textit{red}; 5550 K for MS, \textit{dark blue}) and the corresponding reference colour $(\gk)_0$ (2.49 for RG, \textit{pink}; 1.49 for MS, \textit{light blue}). Note that some lines may overlap. Their cut corresponds to the interval of absorption $A_0$ indicated in Table \ref{tab:empiricalk} (13.3 for RG, 3.5 for MS). }
\label{fig:results}
\end{figure*}

\section{Method} 
\label{sec:method}

To empirically measure the $G$-band extinction coefficient as a function of either temperature or colour ($\gk)_\mathrm{0}$ we implemented 
a Markov-Chain Monte Carlo method (MCMC, \citealt{brooks2011}) to sample the  parameter space  and to properly account for errors.
The MCMC used the jags algorithm \citep{jags} encompassed in \texttt{runjags}
\footnote{\url{https://cran.r-project.org/web/packages/runjags/ runjags.pdf}} library, for R programme language.

In order to not affect the MCMC convergence by having an un-even distribution in extinction and temperature, we used a two dimensional kernel density estimation of the $\Egk$ vs $\teff$ stellar probability space to 
select a more uniform sub-sample of 1000 stars for each RG and MS (Fig. \ref{fig:Data}) 
and combined (RG+MS) sample. 
The number of stars in each sub-sample (i.e. 1000) is optimised to be statistically relevant for the analysis yet without having a large disproportion of elements between bins (which could cause the analysis to be biased towards the most populated bin). 

Intrinsic colours ($\gk)_\mathrm{0}$  and $(\jk)_\mathrm{0}$ 
 were taken from Eq. (\ref{eq:calibration}) where temperature and metallicity were set 
 to be $\teffprime \sim \mathcal{N}(\teff, \sigma_\mathrm{\teff}^2)$ and 
$\feh^{\prime} \sim \mathcal{N}($\feh$, \sigma_\mathrm{\feh}^2)$ where $\mathcal{N}$ is the normal distribution and $\sigma_\mathrm{Teff}^2$, $\sigma_\mathrm{\feh}^2$ the respective observed variances.
 
\noindent Observed colours ($\gk)_\mathrm{obs}$ and ($\jk)_\mathrm{obs}$ were set to be
~
\begin{equation}
\begin{array}{ccl}
(\gk)_\mathrm{obs} &\sim & \mathcal{N}((\gk)_\mathrm{0} + (\mathrm{k}_{G} - \mathrm{k}_{\ks}) \cdot A_0^{\prime} , ~\sigma^2_{G-\ks})\\
(\jk)_\mathrm{obs} & \sim &\mathcal{N}((\jk)_\mathrm{0} + (\mathrm{k}_{J} - \mathrm{k}_{\ks}) \cdot A_0^{\prime} , ~\sigma^2_{J-\ks})
\end{array}
\end{equation}

\noindent where $\sigma^2_{G-\ks} = (\sigma_{G}^2 + \sigma_{\ks}^2$) and $\sigma^2_{\jk} = 
(\sigma_{J}^2 + \sigma_{\ks}^2$) and where $\sigma_{G}, \sigma_{J}, \sigma_{\ks}$
are the photometric errors. 
k$_{J}$ and k$_{\ks}$ are the extinction coefficients for $J$- and $\ks$-bands as function of either $\tnorm$ or colour. All along our analysis k$_{J}$ and k$_{\ks}$ are fixed to the theoretical values (see \S \ref{sec:ASPOS}). \\
For each star in the sample we used its colour excess $\Ejk$ and initial extinction coefficients values (computed at $A_0$ = 0~mag), to get an initial value of the absorption $A_0$, which we then set in the MCMC as mean of a truncated normal distribution $A_0^{\prime}\sim$ $\mathcal{N}(A_0$, 0.2) lying within the positive interval $A_0^{\prime}$ \textgreater ~0.
For a given star, its initial $A_0$ value does not change within the MCMC.
Finally we set the  coefficients $a_{i}$ of Eq. (\ref{eq:coefformula}) free to vary following the uninformative prior distribution $a_{i} \sim \mathcal{N}$(0, 1000).

Each MCMC was run using two chains with $10^4$ steps and a burn-in of 4000. 
We used standardised variables to improve the efficiency of MCMC sampling (hence reducing the autocorrelation in the chains) and checked for chain convergence by 
using the Gelman-Rubin convergence diagnostic. We tested the  significance of coefficients $a_{i}$ through the Deviance Information Criterion (DIC), a model fit measure that penalises model complexity. 

We produced 10 different sub-samples of 1000 stars (uniform in $\teff$ and $\Egk$), each of which was processed through an MCMC. The mean of those runs is reported in Table \ref{tab:empiricalk}.

We run this analysis for the red giants first, then for the main sequence stars, and finally for both samples combined in a single one.

\begin{figure*}[b!]
\centering
\includegraphics[width=0.98\linewidth, trim = 30 0 30 0]{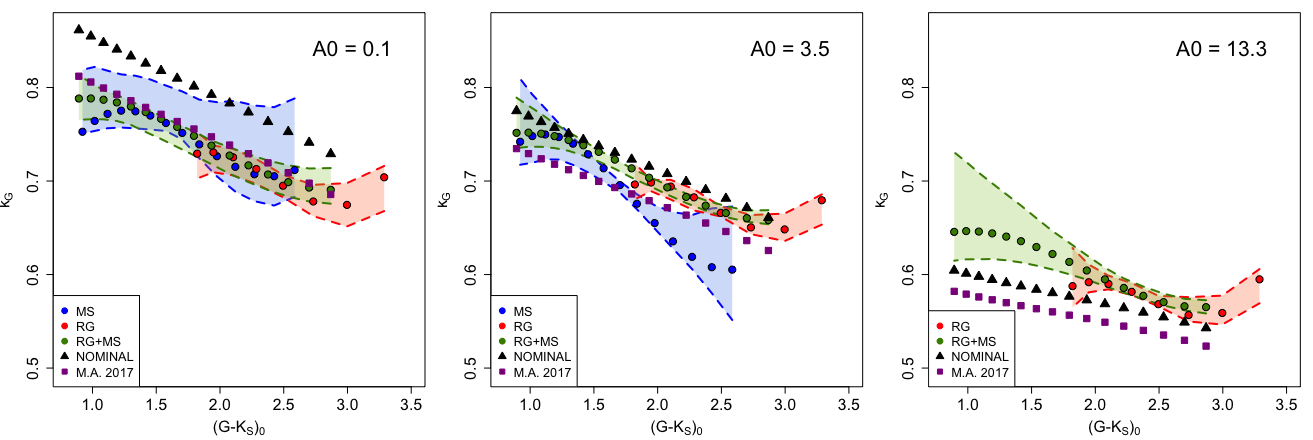}
\medskip
\includegraphics[width=0.98\linewidth, trim = 30 0 30 0]{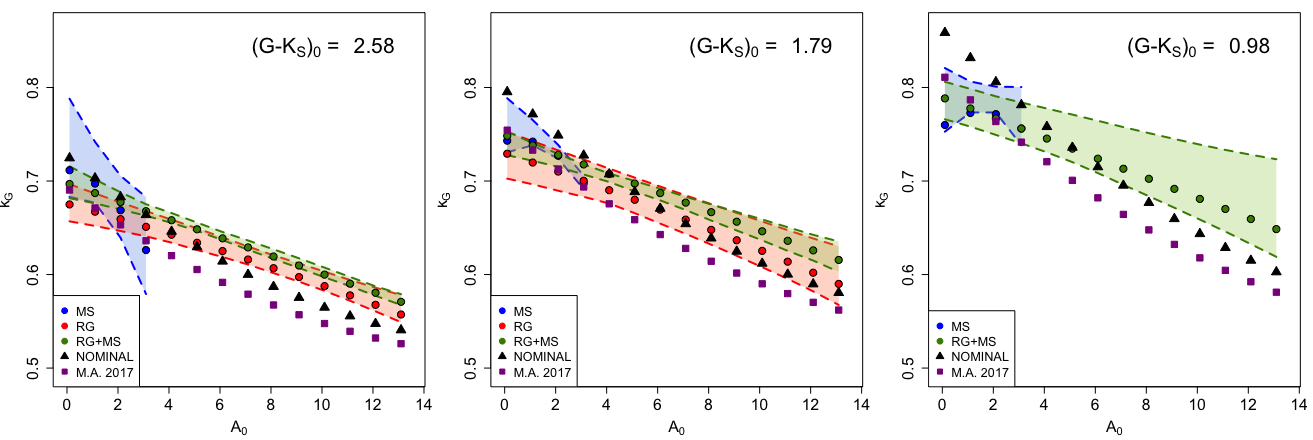}
\caption{Direct comparison of the empirical extinction coefficient k$_{G}$ as a function of ($\gk)_0$
for $A_{0}$ = 0.1, 3.5, 13.3 mag (\textit{top panel}) and $A_0$  for ($\gk)_0$ = 2.58, 1.79, 0.98 (\textit{bottom panel}, which corresponds to $\teff$ = 4020, 5080, 6500 K).
$A_{0}$ = 3.5, 13.3  mag are the limit values for the MS and RG samples respectively (see Table \ref{tab:empiricalk}) while ($\gk)_0$ = 2.58, 1.79 are the maximal colour of LAMOST and the minimal colour of APOGEE, respectively. ($\gk)_0$ = 0.98 is a sample case to show the behaviour at low colour indexes (i.e. high temperatures).
Dots show the  k$_{G}$  mean value while the shaded area show the 95$\%$ interval of confidence (see \S \ref{sec:erranalysis}).
Colours correspond to the red giants (RG, \textit{red}), the main sequence (MS, \textit{blue}) and unified sample (RG+MS,~\textit{green}).
Black triangles and magenta squares show the theoretical  k$_{G}$ coefficient computed with the \textit{Gaia} pre-launch (\textit{NOMINAL}, \citealt{Jordi2010}) and \textit{Gaia} $G$-DR1 (\textit{M.A. 2017}, \citealt{Maiz2017}) passbands, respectively.}
\label{fig:resultall}
\end{figure*}

\subsection{Error analysis}
\label{sec:erranalysis}

To derive our confidence interval, we use the bootstrap resampling technique \citep{Efron1987}. The bootstrap resampling consists of generating a large number of data sets, each with an equal amount of points randomly drawn with replacement from the original sample.
It allows us to take into account not only measurement errors but also sampling-induced errors, which are here a relevant factor due to the uneven distribution of stars in temperature and colour excess space. \\
Bootstrapped $a_i$ errors are larger than MCMC chains errors by an average factor of 
5, 3 and 7 for $\teff$ case and 16, 3 and 4 for the ($\gk)_0$
case for RG, MS and RG+MS, respectively. Important to note though that these uncertainties  are constrained by the precision of the data used, more specifically by the error on the temperature, whose 
median  is $\tilde{\sigma}_\mathrm{Teff} \sim$ 69 K for APOGEE data and $\tilde{\sigma}_\mathrm{Teff} \sim$  115 K for LAMOST  data.

We carried out the MCMC runs on 100 bootstrapped samples and derived the confidence levels through the percentile method, which we report in  Table \ref{tab:empiricalk} and Fig. \ref{fig:resultall}.

\begin{table*}[t!]
 \centering
\fontsize{9}{13}\selectfont
\begin{tabular}{l|c|c| ccccccc}
\multicolumn{10}{c}{EMPIRICAL k$_{G}$ VALUES}\\
 \hline \hline 
  RG & $\teff$ int. & $A_0$ & $a_{1}$ & $a_{2}$ & $a_{3}$ & $a_{4} $&  $a_{5} $& $a_{6}$ & $a_{7}$\\
 \hline 
k$_{G}~[\tnorm, A_0]$& [3680, 5080] & < 13.3 & 15.25& -51.059 & 59.12 & -22.57 & 2.41E-03& -1.42E-04 & -1.19E-02 \\
$\sigma_{kG}~[\tnorm, A_0]$ & & & 3.69  & 13.052 &  15.29  & 5.94 &  4.73E-03 &  9.01E-05 &  5.01E-03 \\
\cline{2-10}
& $(\gk)_\mathrm{0}$ int. & $A_0$  &$a_{1}$ & $a_{2}$ & $a_{3}$ & $a_{4} $&  $a_{5} $& $a_{6}$& $a_{7}$\\
\cline{2-10}
k$_{G}~[(\gk)_\mathrm{0}, A_0]$& [1.82, 2.87]& < 13.3 & -0.72  & 1.94  & -0.84 & 0.116  & -1.24E-02  & -1.07E-04 & 1.68E-03 \\
$\sigma_{kG}~[(\gk)_\mathrm{0}, A_0]$& & & 0.93 &  1.13  & 0.45  & 0.059 &  2.83E-03  &  1.23E-04 &  1.14E-03 \\

 \hline \hline  
MS & $\teff$ int. & $A_0$  &$a_{1}$ & $a_{2}$ & $a_{3}$ & $a_{4} $&  $a_{5} $& $a_{6}$& $a_{7}$\\
\hline
 k$_{G}~[\tnorm, A_0]$ & [4020, 6620]  & < 3.5&  4.40 & -11.405 & 11.52 & -3.77 & -0.051 & -6.60E-03 & 0.056 \\
 $\sigma_{kG}~[\tnorm, A_0]$ & & &  1.38  & 4.028  & 3.85  & 1.21  & 0.044 &  1.65E-03  & 0.038 \\
\cline{2-10}
& $(\gk)_\mathrm{0}$ int. & $A_0$  &$a_{1}$ & $a_{2}$ & $a_{3}$ & $a_{4} $&  $a_{5}$ & $a_{6}$& $a_{7}$\\
\cline{2-10}
k$_{G}~[(\gk)_\mathrm{0}, A_0]$&  [0.92, 2.59] & < 3.5&  0.32&  0.88 & -0.53 &  0.097 &  0.038  & -7.06E-03 & -0.017 \\
$\sigma_{kG}~[(\gk)_\mathrm{0}, A_0]$ & & & 0.15& 0.30 &0.19 &0.036  &0.020 & 1.72E-03 & 0.013  \\

 \hline \hline 
 RG + MS & $\teff$ int.  & $A_0$ & $a_{1}$ & $a_{2}$ & $a_{3}$ & $a_{4} $&  $a_{5} $& $a_{6}$& $a_{7}$\\
 \hline 
 k$_{G}~[\tnorm, A_0]$ & [3680, 6620]& < 13.3 & 3.24  &-8.31  & 8.72  &-2.92  & -7.55E-03 & -8.35E-05 & -6.73E-04\\
 $\sigma_{kG}~[\tnorm, A_0]$ & & & 0.55 & 1.67 & 1.67 & 0.55 & 6.24E-03 & 1.19E-04 & 6.73E-03\\

\cline{2-10}
& $(\gk)_\mathrm{0}$ int. & $A_0$  &$a_{1}$ & $a_{2}$ & $a_{3}$ & $a_{4} $&  $a_{5} $& $a_{6}$& $a_{7}$\\
\cline{2-10}
k$_{G}~[(\gk)_\mathrm{0}, A_0]$& [0.92, 2.87]& < 13.3 & 0.697 & 0.219 & -0.154 & 2.69E-02 & -1.14E-02 & -6.84E-07 & 6.58E-04\\
$\sigma_{kG}~[(\gk)_\mathrm{0}, A_0]$& & & 0.059 & 0.083 & 0.037 & 5.51E-03 & 3.43E-03 & 1.38E-04 & 1.30E-03\\

 \hline
  \end{tabular}
 \caption{Empirical  extinction coefficient k$_{G}$ for the $Gaia$ $G$-band as a function of absorption $A_{0}$ and the normalised temperature $\tnorm$ or colour $(\gk)_\mathrm{0}$  for the red giants sample (RG),  the main sequence sample (MS)  and both samples combined in only one (RG + MS). 
For each sample we report temperature, colour and extinction intervals of validity. The errors (1$\sigma$ uncertainties) on the coefficients have been measured with the bootstrap technique.}
 \label{tab:empiricalk}
\end{table*}

\section{Results and Discussion}
\label{sec:results}

All MCMCs to estimate both k$_{G}[\tnorm, A_{0}$] and  k$_{G}[(\gk)_{0}, A_{0}$] were found to converge for all the three samples analysed (RG, MS and RG+MS). 

Table \ref{tab:empiricalk} reports final $a_i$ coefficients and their uncertainties, as well as k$_{G}$ intervals of validity (i.e. temperature, colour and extinction). The temperature interval (and consequently the colour one) is the range common to all the bootstrapped samples employed in our analysis. 
The maximum extinction ($A_0$) depends on the $\Ejk$ data distribution. 
For conservative reasons, as the colour excess distribution for the three samples is right-skewed (i.e. small number of stars with large colour excess), we  set the $A_0$ upper limit by cutting \textit{ad hoc}  the tail of each distribution after a visual inspection, i.e. where we had small gap in the data or where the stars were too few for giving a robust solution.

We note that, while we tested the significance of high order $a_i$ parameters with the DIC test (see \S\ref{sec:method}), some coefficients in Table \ref{tab:empiricalk} appear as non-significant due to bootstrap errors being significantly larger than the MCMC derived ones.

We show in Fig. \ref{fig:results} the retrieved empirical colour excess
E$(\gk)$ versus E($\jk)$ for the three samples. We picked the median of the high-extinction stars' temperature as reference temperature. 
For the MS sample, the median temperature does not change significantly for high-extinction stars while for the RG the high-extinction stars are the coolest as they are intrinsically significantly brighter. \\
\indent The three stellar samples delivered consistent results. We display in Fig. \ref{fig:resultall} the direct comparison of k$_{G}$ as a function of both colour and extinction.
The "wavy" aspect of the top panel is a direct consequence of the third order polynomial used for the modelling, where the need of the high order had been tested by an ANOVA (see \S \ref{sec:ASPOS}).
The polynomial is well behaved in the interior of the fitting regime, but at the edges it generates 
a phenomenon termed \textquotedblleft polynomial wiggle\textquotedblright, whose main consequence is the  lack-of-accuracy given by the large oscillations of the polynomial at both ends. 
For this reason the accuracy is lower at the borders of the temperature and extinction $A_{0}$ intervals of validity (Fig. \ref{fig:resultall}). For the extinction the effect is less prominent as the polynomial is only of degree two in $A_0$. However, its impact is seen in Fig. \ref{fig:resultall} (top panel, plot 2) where the k$_G$ are not consistent with RG or RG+MS due to
this polynomial edge effect and lack of high extinction stars in the MS sample. \\
\indent While there is a small difference between the empirically retrieved and the theoretical extinction coefficient (both nominal and \textit{G}-DR1 \citep{Maiz2017} passbands), the amplitude and the trend of the variation as a function of temperature (or colour) and extinction is similar. 
Our empirical coefficients are, as expected, closer to the G-DR1 passband than the nominal passband in the low extinction regime. 
However they are larger than the theoretical ones for $A_0$\textgreater 3 mag for the nominal passband, and $A_0 \gtrsim$ 2 mag for the \textit{G}-DR1 passband. 
With the information currently in our possession we are not able to address this issue, which may be due to uncertainties in the extinction law or in the filter response determination, we will though perform the same study for the coming $Gaia$ DR2 release in order to determine the DR2 k$_G$ extinction coefficients and to clarify this problem.


We overall recommend the use of the combined sample (RG+MS) coefficients using the intrinsic colour ($\gk)_0$. The use of the combined sample gives an unique solution for both stellar evolution stages which it is less affected by the polynomial wiggle effect. The colour is 
also less affected by the temperature scale difference between LAMOST and APOGEE.

\section{Conclusions}
\label{sec:conclusions}
We present here the empirical estimation of the $Gaia$ $G$-band extinction coefficient k$_{G}$ that can be used as unique solution for both red giants and main sequence stars.

We used high quality photometry in the $Gaia ~G$-DR1 and 2MASS $J$- and $\ks$-bands combined with the APOGEE DR13 and the LAMOST DR2 spectroscopic surveys to retrieve effective temperatures for red giants and dwarfs samples respectively.
We implemented a Markov Chain Monte Carlo (MCMC) method where we used the photometric calibration $\teff$ vs $(\gk)_\mathrm{0}$ and $(\jk)_\mathrm{0} $ vs $(\gk)_\mathrm{0}$
relations (method presented by \citealt{Laura17}), to estimate the extinction coefficient k$_{G}$ as a function of the normalised temperature $\tnorm$ = $\teff$/5040 or colour $(\gk)_\mathrm{0}$ and absorption $ A_{0}$. 

We compared each empirical k$_{G}$ coefficient (measured for the dwarfs, the red giants and the combined sample) first  with the theoretical one (estimated using both the $Gaia$ $G$-passband modelled pre-launch and the $G$-DR1 \citep{Maiz2017} passband), then between themselves.
For the first case, while we find a small difference between our results and the theoretical 
extinction coefficients for large extinctions, we confirmed that both theoretical and empirical k$_{G}$ have the same trend.
For the second case we find consistent results. 

We modelled the extinction coefficient as a function of both stellar temperature (or intrinsic colour) and absorption to more precisely account for the degeneracy between extinction and spectral energy distribution. We believe that this approach is the best practice, particularly  for large 
passbands such as the $Gaia~ G$-band, where the extinction coefficient varies strongly within the band itself. 

The results presented here are  valid for the $Gaia$ $G$-DR1 band data and they are constrained by the precision of the spectrometric data used for our analysis. 
The same study will be performed for the $Gaia$ DR2 release (April 2018) with the inclusion of the estimation of the extinction coefficient valid for both BP and RP bands.

 \begin{acknowledgement}
 This work was supported by the Centre National d'etudes Spatiales (CNES) post-doctoral funding project.
 This work has made use of data from the European Space Agency (ESA)
mission {\it $Gaia$} (\url{https://www.cosmos.esa.int/gaia}), processed by
the {\it $Gaia$} Data Processing and Analysis Consortium (DPAC,
\url{https://www.cosmos.esa.int/web/gaia/dpac/consortium}). Funding
for the DPAC has been provided by national institutions, in particular
the institutions participating in the {\it $Gaia$} Multilateral Agreement.
 This publication makes use of data products from the Two Micron All Sky Survey, which is a joint project of the University of Massachusetts and the Infrared Processing and Analysis Center/California Institute of Technology, funded by the National Aeronautics and Space Administration and the National Science Foundation.
 \end{acknowledgement}



\bibliographystyle{aa} 
\bibliography{Danielski_Gaia_extcoeff} 

\end{document}